\documentstyle[prl,aps,twocolumn,epsfig]{revtex}
\newcommand{\A}{{\bf A}}

\newcommand{\D}{{\rm d}}
\newcommand{\x}{{\bf x}}
\newcommand{\n}{{\bf n}}
\newcommand{\y}{{\bf y}}

\newcommand{\siml}{\raisebox{-.6ex}{$\stackrel{<}{\scriptstyle{\sim}}$}}
\newcommand{\simg}{\raisebox{-.6ex}{$\stackrel{>}{\scriptstyle{\sim}}$}}

\begin{document}
\twocolumn[\hsize\textwidth\columnwidth\hsize\csname@twocolumnfalse\endcsname
\title{Quantum Game Theory}
\author{Michael L\"assig}
\address{Institut f\"ur theoretische Physik,
Universit\"at zu K\"oln, Z\"ulpicher Str. 77, 50937 K\"oln,
Germany}

\maketitle

\begin{abstract}
A  systematic theory is introduced that
describes stochastic effects in game theory. In a biological
context, such effects are relevant for the evolution 
of finite populations with frequency-dependent selection.
They are characterized by {\em quantum Nash equilibria},
a generalization of the well-known Nash equilibrium points
in classical game theory. The implications of this theory 
for biological systems are discussed in detail.

\vspace{24pt}
\end{abstract}

\vfill
]
\narrowtext
\section{Introduction}

Classical game theory is a well-known mathematical
formalization of competitions with rational rules
and rational players~\cite{game}. The strength of this theory
lies in the abstraction from the detailed scenario.
In its simplest form, a game is reduced to a set of 
basic {\em strategies}
$i = 1,\dots,s$ and a matrix $\A = (A_{ij})$ whose elements
denote the {\em payoff} or relative success of strategy
$i$ played against strategy $j$. A {\em mixed strategy}
$\x = (x_1,\dots,x_s)$ is defined to be a probability 
distribution over basic strategies (i.e., $x_i \geq 0$
for all $i=1, \dots,s$ and $\sum_i x_i = 1$). Rational
playing of mixed strategies will lead to a {\em Nash
equilibrium}, that is, to a strategy $\x^*$ that maximizes
the payoff against itself~\cite{Nash}. This is the central concept 
of classical game theory. A more precise formulation 
will be given below.

Game theory has been widely applied to 
explain the distribution of different phenotypes in biological
populations. These applications are based on a dynamical 
extension called {\em evolutionary game
theory}, which is due to Maynard Smith~\cite{MaynardSmith}. In 
the simplest case, the different phenotypes in a population
are associated with the basic strategies of a game. The 
time-dependent phenotype frequency distribution is then 
a mixed strategy $\x(t)$. It is assumed that the phenotypes
are hereditary and are preserved under the reproduction
process. The payoff of a basic strategy $i$ played against
the population strategy $\x(t)$ enters the 
{\em fitness} $f_i$ of the corresponding phenotype,
that is, the expected number of viable offspring per
individual and per unit of time. 
This game-based population dynamics can be written 
as an evolution equation for the frequency distribution
$\x(t)$. It can be shown that 
{\em every stable fixed point of this dynamics is a Nash 
equilibrium}~\cite{HS}. This result is conceptually important since
it shows how strategic optimization is reached in biological
systems through reproductive success alone, without the 
need for rational thinking. 

Classical game theory is a deterministic theory. A Nash equilibrium
is the outcome of fully rational playing, without any 
effects of chance. Biological game dynamics describes
an equally deterministic course of evolution. Formally,
this corresponds to the limit of infinite populations
obtained by identifying the reproduction rate of a 
phenotype $i$ with its expectation value~$f_i$.

In this paper, we extend game theory to a stochastic
theory including the effects of fluctuations. This
is based on a probabilistic  
game dynamics suitable to describe 
the evolution of finite biological populations. 
A different form of stochastic game dynamics has 
been  been discussed by Marsili and Zhang
in the context of economical systems~\cite{Zhang}. 

In biological populations, fluctuations arise since the actual 
number of viable offspring of a given individual in a given 
time interval differs from its expectation value
determined by the fitness of the individual's phenotype.
This may be caused by a number of different biological
reasons. It will become clear that  
mechanisms producing frequency-dependent fitness values
-- which are the subject of evolutionary game theory --
also give rise to frequency-dependent fluctuations 
and thereby determine the probabilistic dynamics. In this
formalism, the time-dependent population state  
defines a probability distribution over mixed 
strategies, $p(\x,t)$, and evolution takes 
the form of a Schr\"odinger equation in imaginary 
time, $ \partial_t \, p(\x,t) = H p(\x,t)$.
The analogy with quantum physics allows for
a systematic inclusion of fluctuations and 
suggests the name {\em quantum game theory}.

Stochastic effects have been studied extensively
in Kimura and Ohta's theory of {\em neutral evolution}, which
describes the dynamics of populations whose phenotypes
have little or no fitness difference~\cite{Kimura,Ohta}. Fluctuations
are then the dominant force of evolution. In 
particular, it has been shown that an 
initially small fraction  $x_m(t)$
of mutants in an otherwise homogeneous resident 
population can be driven either to extinction or 
to fixation, i.e., $x_m(t) = 0$ or $x_m(t) = 1$
for late $t$. The probabilities for these processes depend 
on the fitness gap between mutants and residents
and on the overall population size, and they determine
the rate of evolution of the population as a whole.  

Fluctuation effects in game theory turn out to be
more involved. Quantum game theory shows a nontrivial interplay 
between deterministic fitness and stochastic forces,
both of which depend on the phenotype frequencies $\x(t)$. 
Hence, a stable stationary state $p^*(\x)$
-- called a {\em quantum Nash equilibrium} --
differs from its classical counterpart. It depends strongly
on two parameters defined below, the characteristic population 
size $n^*$ and the game coupling $\lambda$. We find
a crossover between neutral evolution 
for $\lambda n^* \ll 1$ and classical game theory 
for $\lambda n^* \gg 1$. Fluctuations
can lead to extinction or fixation; the
corresponding probabilities now depend not only on
the population size but also on 
the phenotype considered. We discuss this in detail 
for the simplest game with a mixed classical Nash equilibrium,
the so-called {\em hawk-dove game}.

The paper is organized as follows. In the next section,
we define a game-based classical dynamics suitable 
for finite populations. In section~3, we discuss the 
`quantization' of this dynamics. Section~4 contains a 
detailed analysis of the quantum theory for the hawk-dove
game, and the results are discussed in section~5.

\section{Classical Game Theory and Population Dynamics}

We start by recalling a few well-known definitions and
results of classical game theory. 
In a game with basic strategies $i = 1, \dots, s$ and  
relative payoffs $A_{ij}$, the payoff of
a mixed strategy $\x$ played against another mixed strategy
$\x'$ is taken to be bilinear,  
$\sum_{i,j =1}^s x_i A_{ij} x'_j \equiv \x \A \x' $. 
We can now define a (symmetrical) Nash equilibrium $\x^*$ 
as an optimal strategy against itself, i.e., 
\begin{equation}
\x^* \A \x^* \geq \x A \x^*
\;\;\; \mbox{for all strategies $\x$.} 
\end{equation} 
Consider a population that contains the phenotypes
$i =1, \dots,s$ with time-dependent population numbers
$\n(t) = (n_1(t), \dots, n_s(t))$ and has total size
$n(t) = \sum_i n_i(t)$; these numbers are positive integers.
The evolution of this system can be described
deterministically by an equation of the form 
\begin{equation}
\frac{1}{n_i} \frac{\D n_i}{\D t} = f_i (\n) \;;
\label{nidot}
\end{equation}
the r.h.s.~is called the fitness of the phenotype $i$.
We assume its frequency dependent part is proportional
to the payoff of the basic strategy $i$ played against 
the mixed strategy $\x \equiv \n/n$ in a game with payoff
matrix $\A$, i.e., $f_i (\n) = \lambda (\A \x)_i + \lambda'$,
where $\lambda$ and $\lambda'$ are coefficients independent of $\x$.
The phenotype frequency distribution then obeys the
closed evolution equation  
\begin{equation}
\frac{1}{x_i} \frac{\D x_i}{\D t} = \lambda f_i^{\rm game} (\x)  
\label{xidot}
\end{equation}
with 
\begin{equation}
f_i^{\rm game} (\x) = (\A \x)_i - \x \A \x \;,
\label{fi}
\end{equation}
which is well known in evolutionary game theory~\cite{HS}.
The coupling constant $\lambda < 1$ describes the strength of 
game-based contributions to the fitness. Classically,
it determines only the time scale in eq.~(\ref{xidot})
but it will have a crucial role in the quantum theory.  
The second term on the r.h.s.~of eq.~(\ref{fi}) 
ensures probability conservation, i.e., $\sum_i \D x_i/\D t =0$.

This frequency dynamics is independent of population 
size. In order to include fluctuations, which depend
on absolute population numbers, we start from the 
full dynamics (\ref{nidot}). We choose
\begin{equation}
f_i(\n) = \lambda f_i^{\rm game} (\x) + f^{\rm size} (n) 
\end{equation}
so that the frequencies $x_i(t)$ follow the evolutionary
game dynamics (\ref{xidot}) and decouple from the dynamics
of the total population size,
\begin{equation}
\frac{1}{n} \frac{\D n}{\D t} = f^{\rm size} (n) \;.
\end{equation}
With 
\begin{equation}
f^{\rm size} (n) = \frac{n^* - n}{n^*} \;,
\end{equation}
the latter describes standard logistic growth to 
a stable population size $n^*$. 

This population dynamics is similar to the well-known 
Lotka-Volterra equations. It can be rewritten 
in terms of the scaled population sizes
$\y = (y_1, \dots, y_s) = \n / n^*$,
\begin{equation}
\frac{1}{y_i}\frac{\D y_i}{\D t} 
   = \lambda f_i^{\rm game} (\y/y) + f^{\rm size} (y)
\label{yidot}
\end{equation}
with $y = \sum_i y_i = n/n^*$. For every symmetric Nash 
equilibrium $\x^*$ of the underlying game, eq. (\ref{yidot})
has a stable fixed point $\y^* = \x^*$.

\section{Quantum Game Dynamics}

As discussed above, the actual number of viable 
offspring produced by an individual of phenotype 
$i$ in a given time interval is an integer which 
fluctuates around its expectation value given by
the fitness $f_i$. In any finite population, these
fluctuations produce deviations from the deterministic
dynamics (\ref{nidot}) and lead us to a stochastic 
description of evolution. The population state then
becomes a propability distribution $P(\n,t)$. 
It is convenient to write this in the form of a
quantum state, 
\begin{equation}
| P(t) \rangle = \sum_\n P(\n,t) | \n \rangle
\end{equation}
with 
$| \n \rangle = a_1^{\dagger n_1} \dots 
               a_s^{\dagger n_s} | 0 \rangle $.

The dynamics takes the form of an (imaginary-time) 
Schr\"odinger equation,
\begin{equation}
\frac{\D}{\D t} \, |P(t)\rangle = H |P(t)\rangle \;.
\label{Pdot}
\end{equation}
The Hamilton operator $H$ contains the creation
and annihilation operators $a_i^\dagger$ and $a_i$,
which obey canonical commutation relations and
describe birth and death of an individual of phenotype
$i$, respectively. This formalism has been widely applied
to related dynamical problems such as 
reaction-diffusion models; see, e.g., refs.~\cite{Peliti,Cardy}.

In order to construct the Hamiltonian in a systematic
way, recall the biological meaning of fitness in the
classical theory (\ref{yidot}). The coefficient
function $f_i(\y)$ is not a simple birth rate but
the effective rate of reproductive success, taking
into account birth and death processes. It can be
written as the difference between a birth rate $b_i(\y)$
and a death rate $d_i(\y)$ which are both positive,
\begin{eqnarray}
f_i(\y) & =  b_i(\y) - d_i(\y) 
              =  & \lambda (b_i^{\rm game} (\x) 
	      - d_i^{\rm game} (\x))
\nonumber \\
        &   & + b^{\rm size} (n) - d^{\rm size} (n) \;.  
\label{decomp}
\end{eqnarray}

Clearly, this decomposition requires additional biological 
input as will be discussed below for the hawk-dove game. 
In the quantum theory, it 
becomes important since birth and death processes have
individual fluctuations. As operators, these rates determine
the reproduction current, 
\begin{equation}
J_i =   a_i^\dagger a_i b_i (\hat \n) - a_i d_i (\hat \n) \;.
\label{Ji}
\end{equation}  
with $\hat \n = (\hat n_1, \dots, \hat n_s) = 
                (a_1^\dagger a_1, \dots, a_s^\dagger a_s)$.   
We have 
$J_i |P(t)\rangle = \sum_\n j_i (\n,t) |\n \rangle$,
where 
\begin{equation}
j_i (\n,t) = n_i b_i(\n) P(\n,t) - 
          (n_i +1) d_i(\n + i) P(\n + i,t)
\label{jn}
\end{equation}	  
is the probability current between the states $|\n \rangle$	  
and $|\n + i\rangle$. Here we use the shorthand
\begin{equation}
\n \pm i  \equiv 
   (n_1,\dots,n_{i-1},n_i \pm 1,n_{i+1},\dots,n_s) \;. 
\end{equation}   
The Hamiltonian then takes the form 
\begin{equation}
H  =   \sum_{i=1}^{s} (a_i^* -1) J_i 
\label{H}
\end{equation}
so that eq.~(\ref{Pdot}) is  equivalent
to the Master equation
\begin{eqnarray}
(\partial/\partial t) \, P(\n,t) 
  & = & (n_i -1) b_i (\n - i)  P (\n - i,t)  \nonumber \\
  &   & -n_i     b_i (\n)      P (\n,t)       \nonumber \\ 	   
  &   & -n_i     d_i (\n)      P (\n,t)       \nonumber \\
  &   & +(n_i +1) d_i (\n + i)  P (\n + i,t)   \;.
\label{master}
\end{eqnarray}

Probability conservation implies that $H$ only has
eigenvalues with non-positive real parts. We are 
interested in particular in the leading eigenstate
$|\Phi^* (t) \rangle$ with support in the coexistence 
region of all phenotypes, 
$n_1 \geq 1, \dots, n_s \geq 1$. 
This eigenstate decays due to extinction processes
$n_i =1 \to 0$. The eigenvalue $E^* < 0$ can be written
in terms of the extinction current given by eq.~(\ref{jn}),
\begin{equation}
\frac{E^*}{{\cal N}(t)} =    
      \sum_{i=1}^s \sum_{\n |n_i =0} j_i^*(\n,t) =
      \sum_{i=1}^s \sum_{\n |n_i =1} d_i(\n) \Phi^*(\n,t)
\label{Estar}
\end{equation}
with ${\cal N}(t) =  
      \sum_{n_1 \geq 1,\dots,n_s \geq 1} \Phi^* (\n,t)$.

For further analysis, let us approximate the Master
equation (\ref{master}) by a continuous diffusion
equation, which is conveniently written for a population
state depending on the scaled variables $\y$,
\begin{eqnarray}
\frac{\partial}{\partial t} \, \Phi(\y,t)
 & = & \frac{1}{2 n^*} 
  \sum_{i=1}^s \frac{\partial^2}{\partial y_i^2} \,
                  G_i(\y) \Phi(\y,t) \nonumber \\
 &   & - \lambda \sum_{i=1}^s \frac{\partial}{\partial y_i} \,
		  V_i(\y) \Phi(\y,t) 
\end{eqnarray}
with 
\begin{eqnarray}
G_i(\y) & = &  y_i [b_i (\y) + d_i(\y)] \;, \\
\lambda V_i(\y) & = &  y_i [b_i (\y) - d_i(\y)] \;.
\end{eqnarray}

Biological populations may be sufficiently large
so that their variation in relative size, which 
is of order
\begin{equation}
\langle (y-1)^2 \rangle \sim 1/n^* \;,
\end{equation} 
can be neglected. The fluctuations in phenotype composition 
may still be significant, depending on the scaled game coupling
constant $\lambda n^*$. They are described by a projected
diffusion equation for the phenotype state
\begin{equation}
\phi(\x,t) \equiv \int \D y \, \Phi(\y,t) \delta(\x - \y /y) \;.
\label{pro}
\end{equation}
Long-term fluctuations are characterized by the leading 
eigenstate $\phi^* (\x,t)$ with support in the coexistence 
region $0 < x_i < 1$  
and the corresponding eigenvalue $e^* < 0$. Removing the
time-dependence by normalization defines
a stationary phenotype probability distribution $p^*(\x)$,
the quantum Nash equilibrium. For game-based
evolution, the diffusion coefficients 
and the drift fields have a nontrivial
dependence on the phenotype composition $\x$, preventing
a solution in closed form. 
Therefore, we will now turn to a specific example.

\section{The Quantum Hawk-Dove Game}

The hawk-dove game~\cite{MaynardSmith} 
is one of the simplest classical
games. `Hawks' ($i=1$) and `doves' ($i=2$) are the two 
phenotypes of this game, and there is a single independent
frequency variable $x \equiv x_1 = 1 - x_2$. The game is 
defined by the (suitably normalized) payoff matrix
\begin{equation}
\A = \left ( \begin{array}{cc} 1-c & 2 \\ 0 & 1 
           \end{array} \right )
\end{equation}
with a constant $c > 1$, leading to a classical population
dynamics of the form (\ref{xidot}), (\ref{fi}), 
\begin{equation}
\frac{{\rm d} x}{{\rm d} t } = \lambda x (1-x) (1 - c x) \;.
\end{equation}
The unique stable fixed point is the  
Nash equilibrium $x^* = 1/c$. 

The hawk-dove game dynamics has been used to explain 
instinctive agression control in animal species. Individuals
of the same species may fight for territory, mating
partners, etc. `Hawks' are aggressive individuals who 
escalate fights, while `doves' avoid escalation and retreat.
Hence, a hawk will always win against a dove ($A_{12} = 2,
A_{21} = 0$), compared to the payoff of doves againts doves
($A_{22} = 1$). However, hawk-hawk encounters involve a
fitness decrease due to mutually inflicted injuries ($A_{11}
= 1 -c$). In species with heavy weaponry ($c \gg 1$), 
fights are usually
limited to ritual display of force and truly
aggressive individuals are rare, as indicated by 
the Nash equilibrium $x^* \ll 1$. Needless to say, this
is but the simplest approximation to the complexities
of animal behavior. 

The biological interpretation also suggests a
decomposition (\ref{decomp}) of the fitness into birth and
death processes. Here we choose
\begin{equation}
\begin{array}{lll}
b_i^{\rm game} (\x) & = & ({\bf B} \x)_i - \x {\bf D} \x \;, 
\\
d_i^{\rm game} (\x) & = & ({\bf D} \x)_i - \x {\bf B} \x 
\end{array}
\label{bidi}
\end{equation}
with 
\begin{equation}
{\bf B} = \left (\begin{array}{cc} 1 & 2 \\ 0 & 1 
           \end{array} \right ) \;,
\;\;\;
{\bf D} = \left (\begin{array}{cc} c & 0 \\ 0 & 0 
           \end{array} \right ) \;.
\label{BD}	   
\end{equation}
The matrix ${\bf B}$ describes game-related variations of
the birth rates, while ${\bf D}$ contains the deaths due
to fights. The offset terms $\x {\bf B} \x$ and $\x {\bf D}
\x$ required by eq.~(\ref{fi}) are associated with the other
type of process, respectively. The population size is
assumed to be controlled by a bare birth rate and a variable
death rate,
\begin{equation}
b^{\rm size} = 1 \;,
\;\;\;
d^{\rm size} = y \;.
\label{bdsize}
\end{equation}
Of course, this decomposition is not unique, and a different
biological context may suggest a different choice. The
qualitative results described below do not depend on these
details. 

Using the decomposition (\ref{bidi})--(\ref{bdsize}), we 
obtain the one-dimensional phenotype diffusion equation
\begin{equation}
\frac{\partial}{\partial t} \phi(x,t) = 
   \frac{1}{2 n^*} \frac{\partial^2}{\partial x^2} 
   g(x)\phi(x,t) - 
   \lambda \frac{\partial}{\partial x} 
   v(x)\phi(x,t) 
\end{equation}
with    
\begin{eqnarray}
g(x) = & x (1-x) & [ (1-x) (b_1(x) + d_1(x)) 
\nonumber \\
            &    & + x (b_2(x) + d_2(x))]
\end{eqnarray}
and
\begin{equation}
v(x) = x (1-x) (1 - x/x^*) \;.
\end{equation}

The quantum Nash equilibrium $p^*(x)$ and the eigenvalue
$e^*$ can be computed approximately. For $\lambda = 0$,
the solution is $p^*(x) = 1$ and $e^* = 1/2 n^*$, describing
the well-known case of neutral evolution. Corrections 
for $\lambda n^* \ll 1$ have the form of a power series 
in $\lambda n^*$.  
For $\lambda n^* \gg 1$, obtain the approximate functional 
functional form of the solution by neglecting extinctions,
i.e., by setting $e^* \approx 0$. The current in the Nash state,
$j^* (x) \equiv (1/2n^*) (\D /\D x) (g(x) p^*(x)) - \lambda
v(x) p^*(x)$, is also zero in this approximation, yielding
\begin{equation}
p^*(x) = p^*(x^*) \,
         \frac{g(x^*)}{g(x)} \, 
         \exp \left [ 2 \lambda n^* 
                      \int_{x^*}^x \D x' \frac{v(x')}{g(x')}
              \right ] \;.
\label{pxstar}
\end{equation}
For very large populations, $p^*(x)$ is seen to be a
Gaussian of width $\sim 1/\sqrt{\lambda n^*}$ centered 
around the classical Nash equilibrium $x^*$.
The approximation (\ref{pxstar}) is seen to be self-consistent for
\begin{equation}
x_{\min} \simeq \frac{1}{2 n^* \lambda}
           \frac{g'(0)}{v'(0)} 
	\, \siml \, x \, \siml \,
x_{\max} \simeq 1 - \frac{1}{2 n^* \lambda}
           \frac{g'(1)}{v'(1)} \;.
\end{equation}
Of course, the current cannot be neglected close to the boundary,
where we have 
$j^*(x) = (-1/2 n^*) g'(0) \phi(0) + O(x)$ and
$j^*(x) = (-1/2 n^*) g'(1) \phi(1) + O(1-x)$. Matching the
two regimes and using (\ref{Estar}) gives an asymptotic 
expression for $e^*$. The ratio of the extinction 
currents takes the simple form
\begin{equation}
\frac{j^*(1)}{- j^*(0)} \simeq 
   \exp \left [ 2 \lambda n^* \int_{x_{\min}}^{x_{\max}}
                              \D x' \frac{v(x')}{g(x')}
	\right ] \;.
\label{j0j1}
\end{equation}
Fig.~1 shows the quantum Nash state $p^*(x)$ obtained 
numerically for various values of the scaled game coupling
$\lambda n^*$. The solution of the full Master equation
(\ref{master}) is projected onto the frequency variable $x$
using eq.~(\ref{pro}).
The crossover from neutral to game-dominated
behavior is seen already at small equilibrium population
sizes (here $n^* = 70$). 
The strong-coupling approximation
(\ref{pxstar}), also shown in fig.~1, turns out to be an
excellent approximation in the regime $\lambda n^* \simg 1$,
$ x_{\min}(\lambda,n^*) \, \siml \, x \, \siml \, 
  x_{\max}(\lambda,n^*)$. 			 	   

\begin{figure}[t]
\vspace*{-0.1\textheight}
\includegraphics[width=0.45\textwidth]{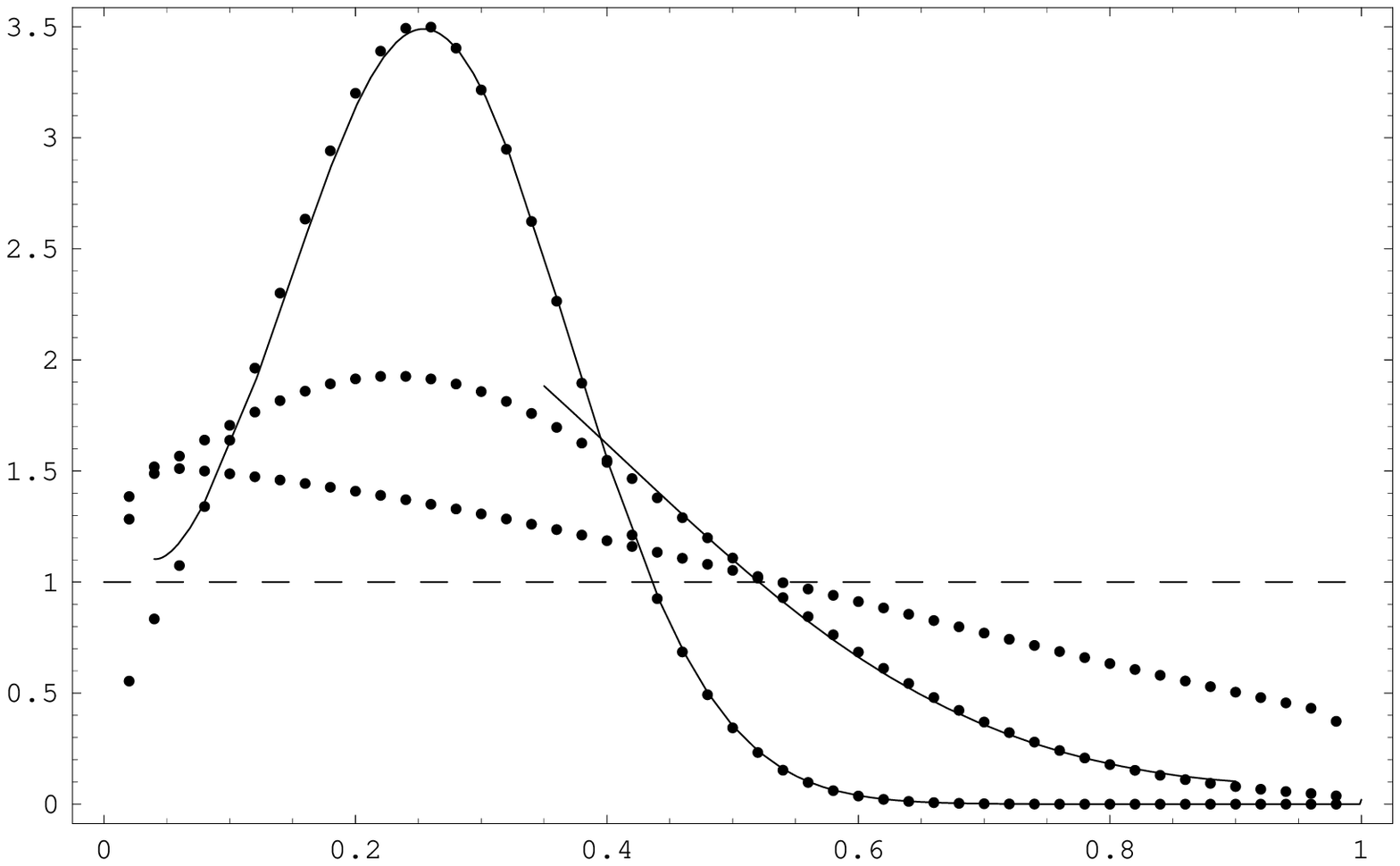}
\vspace*{-0.12\textheight}

\noindent
{\small Fig.~1. The quantum Nash equilibrium $p^*(x)$ for 
$n^*= 70$ and $x^* = 0.28$. 
Numerical solution of the Master equation (\ref{master}) 
for $\lambda=1, 0.1,0.01$ (dots), the limit $\lambda=0$
of neutral evolution (dashed line), and the strong-coupling
solution (\ref{pxstar}) for 
$ x_{\min}(\lambda,n^*) \, \siml \, x \, \siml \, 
  x_{\max}(\lambda,n^*)$
(solid lines).}
\end{figure} 

A comprehensive analytical treatment of the quantum
hawk-dove game is beyond the scope of this paper. 
Instead we now discuss properties of the solution
which are valid beyond this specific example.

\section{Results and Discussion}

The quantum Nash equilibrium $p^*(x)$ describes
the likelihood of finding phenotype frequencies $\x$
in long-term observations of biological sample 
populations. The fluctuations in phenotype composition
depend on the scaled game coupling constant $\lambda n^*$
and may thus be significant even in large populations
where the relative size fluctuations can be neglected. 
There are two modes of evolution. 

In the {\em weak-coupling}
regime $ \lambda n^* \ll 1$, the  system 
is dominated by stochastic forces. These produce a
broad quantum Nash equilibrium,  which implies large
frequency variations,
\begin{equation} 
\langle (\x - \langle \x \rangle )^2 \rangle \sim 1 \;.
\end{equation}
In the limit $\lambda = 0$, the evolution becomes strictly
neutral. 

In the {\em strong-coupling} regime $ \lambda n^* \gg 1$,
the game-related deterministic forces become relevant. 
These systems have only small frequency variations around
a classical Nash equilibrium,
\begin{equation}
\langle (\x - \x^* )^2 \rangle \sim 1/ \lambda n^* \;,
\end{equation}
as shown by the quantum Nash state (\ref{pxstar}) for the
hawk-dove game. Classical game theory is recovered in the
limit $n^* \to \infty$. 

At any finite $\lambda n^*$, the mixed Nash equilibrium can 
be altered drastically by the extinction of phenotypes.
The probability of extinction depends strongly on the 
phenotype considered. In the hawk-dove game, hawks face
a much higher risk of extinction than doves, as indicated
by the ratio of the extinction currents in the quantum 
Nash state given by (\ref{j0j1}). For $x^* < 1/2$, the
current $j^*(0)$ is exponentially larger in magnitude than $j^*(1)$.
This result also implies that an initially small hawk 
mutant population in a dove resident population is 
less likely to grow to its classical equilibrium frequency 
$x^*$ than a dove mutant in a hawk resident population.

In a similar way, stochastic effects may influence the 
internal evolution of phenotypes even in the strong-coupling
regime, where the overall frequencies are close to the
classical Nash equilibrium $\x^*$. In the hawk-dove
game, consider a mutant phenotype with a linkage
disequilibrium such that it can invade only the hawk
subpopulation. According to the standard theory of neutral 
evolution, the fixation probability of this mutant depends
on the fitness difference to the resident hawks and on the
effective hawk population size $n_1^{\rm eff}$. It is easy 
to show that this may be much smaller than the effective
population size $n_2^{\rm eff}$ for doves, as given by the 
ratio
\begin{equation}
\frac{n_1^{\rm eff}}{n_2^{\rm eff}} = 
  \frac{x^*}{1-x^*} 
  \frac{b_2(x^*) + d_2(x^*)}{b_1(x^*) + d_1(x^*)} \;.
\end{equation}
Hence, hawks face a larger mutation load and a lesser
chance of fixation for adaptive mutations than doves.

To summarize:
In classical game theory, the basic strategies represented
in a mixed Nash equilibrium $\x^*$ are equivalent in the 
sense that their payoffs $(\A \x)_i$ are all equal to the
average payoff $\x \A \x$, see eq.~(\ref{xidot}). Stochastic 
effects break this equivalence and alter the equilibrium
state. In biological systems, stochasticity is caused by
frequency-dependent birth and death rates, that is, by
the same mechanism that gives rise to frequency-dependent
fitness values and underlies the application of classical 
game theory. Stochastic evolution creates a bias against 
phenotypes with larger fluctuations in their birth and death rates.   			 

Quantum game theory opens a systematic way to quantify 
these stochastic effects. It contains Maynard-Smith's evolutionary 
game theory and Kimura's theory of neutral evolution
as the limit cases of weak and strong stochasticity,
respectively. We have introduced here the 
relevant concepts and applied them to the simplest kind
of game. Obvious extensions include more complicated games 
such as bimatrix games. 

Another avenue for future work is cooperative
fluctuations in phenotype and {\em space}. In biological
systems, population states $p(\x, {\bf r})$ now depend 
also on spatial coordinates, and stochastic game theory 
becomes a non-equilibrium quantum field theory.  
Population dynamical problems involving diffusion and
migration have been treated by field-theoretic 
methods~\cite{NelsonShnerb}, 
and more recently the process
of biological speciation has been 
identified as a coupled phase separation in phenotype 
and real space~\cite{LassigRost}.

Finally, it is tempting to speculate about a different 
kind of randomness in game dynamics. The stochastic
effects discussed so far originate from events random in 
time. However, we may also consider games with a payoff
random for individual pairings of players but independent 
of time. This {\em quenched disorder} may affect Nash
equilibria in new ways.

\section*{Acknowledgments}
It is a pleasure to thank Peter Hammerstein 
and Antony Zee for inspiring discussions.

\end{document}